**Method to monitor the evolution of an epidemic in real time**

Justin Trujillo [1] and Valerică Raicu [1, 2,*]

[1] *Physics Department, University of Wisconsin-Milwaukee, Milwaukee, Wisconsin 53211, USA*

[2] *Department of Biological Sciences, University of Wisconsin-Milwaukee, Milwaukee, Wisconsin 53211, USA*

The emergence of an epidemic evokes the need to monitor its spread and assess and validate any mitigation measures enacted by governments and administrative bodies in real time. We present here a method to observe and quantify this spread and the response of affected populations and governing bodies and apply it to COVID-19 as a case study. This method provides means to simultaneously track in real time quantities such as the mortality and the recovery rates as well as the number of new infections caused by an infected person. With sufficient data, this method enables thorough monitoring and assessment of an epidemic without assumptions regarding the evolution of the pandemic in the future.

The spread of COVID-19 across the world has prompted the need to further understand the transmission of infectious diseases. This pandemic has displayed distinctive features regarding virulence, spread in real time, and a later onset of symptoms compared to the outbreak of SARS coronavirus almost two decades ago [1]. This has led to a necessity to develop methods with which to draw consistent conclusions about the geographic disparities of the disease [2] as well as provide means to monitor the effects of social interventions aimed at reducing mortality. Furthermore, differences in mortality and recovery rates between affected nations highlight the importance of assessing the self-consistency of reported data for a given nation or municipality, which might suggest ways to improve data collection and reporting. In this work, we draw inspiration from an approach previously introduced to study relaxation in complex systems [3, 4] in order to develop a real-time process by which the evolution of an epidemic may be monitored and apply it to COVID-19 to demonstrate its use. This method allows for the rapid assessment of the effect of mitigation measures (e.g., stay-at-home or lockdown orders) introduced by some countries and, where data is available, by states and local municipalities.

In the current approach, the progression of an epidemic through a population may be described as a system with a self-similar, hierarchical arrangement of coupled states which branch from an initial excited state into a series of subsequent excited states populated by infected people, with a multiplicity or branching factor *b*, and ultimately terminate with states of de-excitation, namely the dark state (death) and the recovered state (see Fig. 1). Each excited (infected) state acts as a "donor" which passes virus to an "acceptor" in direct proportion to the number of individuals in the ground (i.e., uninfected) state, via a rate constant $\Gamma$. The individuals in the second excited state in turn act as donors for additional people in the ground state, and so on. From any of the excited states, the population decays to either the dark (through death) or the recovered state at rates $\delta$ or $\rho$, respectively. Thus, the spread of a disease and the response of the affected population is similar to a relaxation process, where de-

---

excitation from an excited state to the dark or recovered state may be termed *parallel relaxation* [5]. This structure is aptly described as a Cantor-type fractal network which has been used to model dielectric and optical relaxation in complex systems [4, 5], whereby the present system relaxes through a parallel relaxation process (at rates $\delta$ and $\rho$). A notable difference in the present situation versus that of Yokoi et al. [4] and Badu et al. [5] is that there is no *serial relaxation* in an epidemic, as the transmission of the virus from one generation of branches to another (at rate $\Gamma$) does not deplete the population in the previous generation. Additionally, the rate constants $\delta$ and $\rho$ may change in time as the affected population modifies its collective behavior in response to the epidemic.

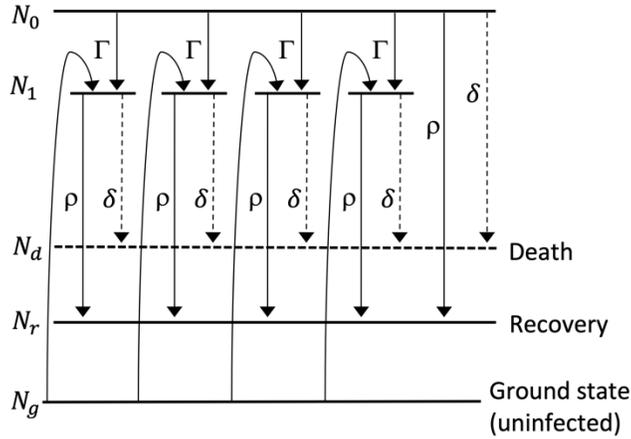

**FIG. 1.** Example of a Cantorian-type fractal relaxation network in which each state branches out evenly with a branching factor (e.g., number of people infected by a single person) *b = 4*. The network shown has a ground state representing the uninfected population, $N_g$, two generations of excited states with $N_0$ and $N_1$, representing the number of people that are infected at a given point in time, a dark state $N_d$ (representing death), and a recovered state, $N_r$. The symbols $\Gamma$, $\delta$, and $\rho$ represent the rate of infection, death and recovery, respectively. Curved arrows represent the transfer of people from the ground state to an excited state (infection).

The present model also resembles a process of Förster resonance energy transfer (FRET) [6], with multiple generations of donors and acceptors and two dark states, one with an infinitely long time constant (corresponding to fluorophore photobleaching) and the other with some finite time constant (triplet state). This similarity suggests a possible way to generalize the present model, by including subsequent excitation "pulses," once the pandemic has led to significant changes in the population residing in the ground state and/or the immunity of the population in the "Recovered" state has waned due to significant mutation of the virus [7] and have thus returned to the ground state. At this still relatively incipient stage of the rapid spread of COVID-19 throughout the world, such refinements could obfuscate the simple meaning of the transmission mechanisms and the mathematical quantities involved and will therefore not be adopted.

The system of equations describing the populations of the various excited-state levels (i.e., individuals becoming infected at different times), $N_i(t)$ with $i = 0, \dots, n$ (see Fig. 1), is

$$\frac{dN_0(t)}{dt} = -N_0(t)\gamma$$

$$b_0 \frac{dN_1(t)}{dt} = b_0 N_0(t)\Gamma - b_0 N_1(t)\gamma$$

$$b_0 b_1 \frac{dN_2(t)}{dt} = b_0 b_1 N_1(t)\Gamma - b_0 b_1 N_2(t)\gamma$$

. (1)

.

.

$$\prod_{i=0}^{n-1} b_i \frac{dN_n(t)}{dt} = \prod_{i=0}^{n-1} b_i N_{n-1}(t)\Gamma - \prod_{i=0}^{n-1} b_i N_n(t)\gamma.$$

where $\gamma = \rho + \delta$ is the rate constant for the change in the number of infections (i.e., the rate at which an outcome is achieved). In the present work, *b* and $\Gamma$ often appear as products with one another. All symbols are defined above and in the caption of Fig. 1.

In addition to equations (1), the rates at which the dark and recovered states are populated may be written as

$$\frac{dN_d(t)}{dt} = N_a(t)\delta, \tag{2}$$

$$\frac{dN_r(t)}{dt} = N_a(t)\rho, \tag{3}$$

where $N_a = N_0(t) + b_0 N_1(t) + b_0 b_1 N_2(t) + \cdots + \prod_{i=0}^{n-1} b_i N_n(t)$ is the number of active cases (i.e., people still infected) at any given time. From Eqs. (2) and (3), we obtain the two rate constants,

$$\delta = \frac{1}{N_a(t)} \frac{dN_d(t)}{dt}, \tag{4}$$

$$\rho = \frac{1}{N_a(t)} \frac{dN_r(t)}{dt}. \tag{5}$$

Usually, one integrates numerically the system of equations (1) to determine the concentrations in each branch of the network as a function of time and to compute, e.g., the cumulative distribution function of "particles" in excited states following excitation, also known as the relaxation function [4, 5]. The relaxation function is neither necessary nor appropriate to use here, since each infected person may act as an additional "excitation," unlike the situation seen in physical relaxation processes. Instead, by integrating Eqs. (2) and (3) from 0 to *t* and assuming that $\rho$ and $\delta$ are independent of time, we obtain:

$$N_d(t) = N_d(0) + \int_0^t \delta(t) N_a(t) dt, \tag{6}$$

$$N_r(t) = N_r(0) + \int_0^t \rho(t) N_a(t) dt. \tag{7}$$

A *cumulative fatality ratio* may then be defined as the ratio between the number of deaths to the number of cases which had an outcome (deaths plus recoveries) recorded from the beginning of the epidemic to the present time, that is,

$$F = \frac{N_d(t)}{N_d(t)+N_r(t)}. \tag{8}$$

This quantity is best used at the end of the epidemic – when all cases would have had resolved either into recovery or death – as it provides an overall picture of how deadly the epidemic has been, regardless of the factors involved (i.e., biological vs. those pertaining to the quality of the healthcare system).

An *instantaneous fatality ratio* also be introduced, namely

$$f = \frac{\frac{dN_d(t)}{dt}}{\frac{dN_d(t)}{dt}+\frac{dN_r(t)}{dt}} = \frac{\delta}{\delta+\rho}, \tag{9}$$

which is sensitive to instantaneous changes in the quantities involved. Provided enough statistics are available, this equation may be used at any time during the outbreak for real-time monitoring of the effects of various measures implemented by policymakers to reduce person-to-person spread of the virus, and to analyze the improvement (or reduction) in the quality of patient care.

Ideally, the cumulative and instantaneous fatality ratios should take identical values. Indeed, using Eqs. (4) and (5), and assuming that $\int_0^t \delta(t)N_a(t)dt \gg N_d(t=0)$, $\int_0^t \rho(t)N_a(t)dt \gg N_r(t=0)$, and that $\delta(t)$ and $\rho(t)$ depend on time much more slowly than $N_a(t)$ [or, equivalently, than $N_0(t)$, $N_1(t),…, N_n(t)$], we see that Eq. (8) takes the approximate form:

$$F \cong \frac{\delta}{\delta+\rho}. \tag{10}$$

Nevertheless, the actual recovery and death rates ($\rho$ and $\delta$, respectively) may sometimes change rapidly because of (i) insufficient resources for quality healthcare, (ii) improvements in those resources, (iii) various measures taken by the relevant authorities for reducing social contact, (iv) the specific way in which citizens respond to those measures, etc. Thus, the approximation for the *cumulative fatality ratio* given by Eq. (10) is rarely valid. Furthermore, if monitoring of infectious cases started after the epidemic was well underway, which is often the case, the actual $N_d(t)$ and $N_r(t)$ remain unknown, making it difficult to compute a cumulative fatality ratio with a high degree of accuracy.

Another useful quantity is the ratio of the number of new infections to the number of cases which newly achieved an outcome (recovery or death), termed the *new infection ratio*,

$$\frac{b_0 \Gamma}{\gamma} = \frac{\frac{dN_T(t)}{dt}}{\left(\frac{dN_d(t)}{dt}+\frac{dN_r(t)}{dt}\right)}, \tag{11}$$

where $N_T(t) = N_a(t) + N_r(t) + N_d(t)$ is the total number of cases at a given time. The complete derivation is found in the Supplemental Material and is briefly summarized here. We first introduce the *transmission rate* (see the Supplemental Material),

$$b_0 \Gamma = \frac{1}{N_a(t)}\frac{dN_T(t)}{dt}. \tag{12}$$

Adding Eqs. (4) and (5) together, we arrive at

$$\gamma = \frac{1}{N_a(t)}\left(\frac{dN_d(t)}{dt} + \frac{dN_r(t)}{dt}\right), \tag{13}$$

which is a rate constant introduced via equation (1). Dividing Eq. (12) by (13), we obtain the new infection ratio, Eq. (11), as desired. The new infection ratio may be interpreted as the average number of infections caused by one infected person at a given point in time.

A quantity similar to that defined by equation (11), known as the reproduction number or $R_0$ value, is often used to quantify the contagiousness of a disease [8]. $R_0$ analyzes population dynamics of susceptible, infected, and recovered (S-I-R) persons and generally relies on unchanging rate constants, a complete knowledge of these populations since the onset of the epidemic [9], as well as inferences about how the epidemic is anticipated to evolve in the future [10]. However, this model is frequently modified with other parameters such as the period of contagiousness and is usually estimated retroactively to gain an overall understanding of the effect of the specific disease on a specific population [11]. What distinguishes the new infection ratio defined in this work from similar quantities (e.g., the $R_0$ value) is that the new infection ratio may be used in conjunction with other elements of the present model to obtain a complete and real-time characterization of an epidemic within a single, unified framework without relying on assumptions of how the epidemic will evolve in the future.

It is useful to monitor if and how the instantaneous fatality ratio given by equation (9) changes as a result of various policies implemented by policymakers in their attempt to reduce the number of infections or the speed with which they change (which would increase exponentially in the absence of any intervention). To observe the fatality ratios and other quantities, data from the COVID-19 Dashboard by Johns Hopkins University was analyzed for some of the countries most affected by the current COVID-19 pandemic [12, 13]. Original data plots are reproduced in the Supplementary Online Materials for reference. Figure 2 displays comparatively the two fatality ratios [Eqs. (8) and (9)] as a function of time computed from the original data for eight countries, and may be applied to other countries (see also Supplemental online figures), states, municipalities, etc. as a tool for monitoring the effect of various administrative and social changes occurring in response to the epidemic, so long as sufficient data is available.

A general trend seen in all plots is a sudden increase followed by a rapid decrease in the early stage of the epidemic. This transient process stems from the nonlinear behavior of the system of equations, while its rather noisy appearance reflects the ramping up of the testing capability of each country, until the tests are able to eventually reflect the real number of cases. The transient occurred more than once in the cases of the USA and Brazil, possibly due to false starts in the testing system or because the epidemic started at different times in different states and so too did the testing process. Beyond such transients, the instantaneous fatality ratio vs. time and other plots provide rich information on the evolution of the epidemic.

While the cumulative fatality ratios present significant inertia which obfuscates any changes caused by various containment measures taken by governmental leadership, the instantaneous fatality ratio more clearly revealed those effects. Following implementation of containment measures and after a brief adjustment period, the fatality ratios begin to decline, as seen in Fig. 2. Notably, South Korea began implementing widespread testing and contact tracing early, so the fatality ratios began to fall before the officially implemented social-distancing measures were enacted and hovered around 2%. Contrarily, introduction of containment measures in France and Brazil did not appear to significantly reduce the instantaneous fatality ratio, which continued to rise before settling to about 10%.

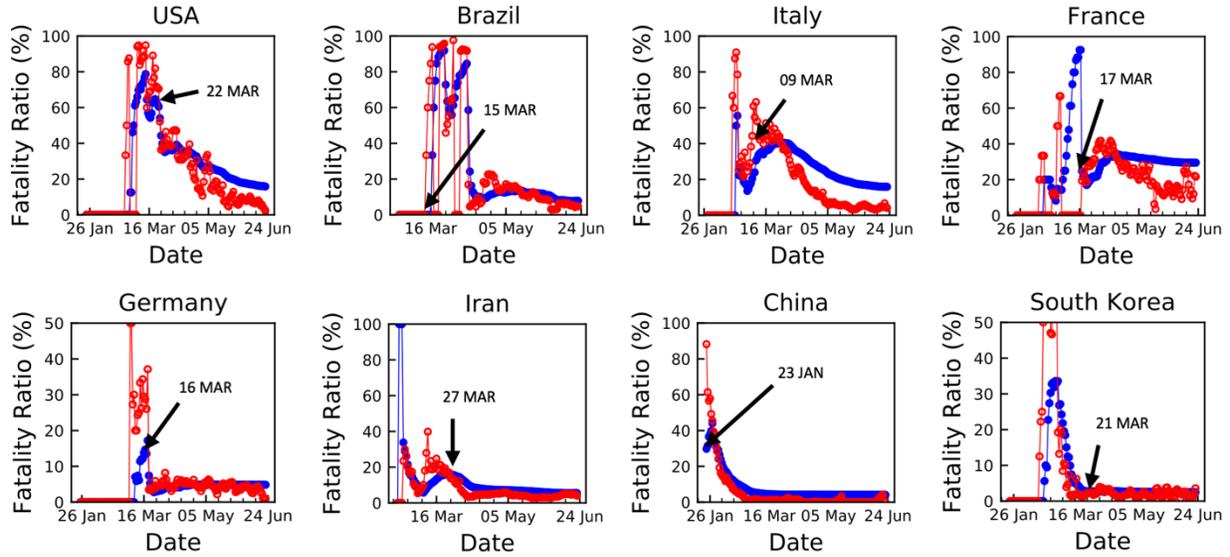

**FIG. 2.** Comparison between the cumulative fatality ratio $F$ [Eq. (8)] (filled blue circles) and the real-time fatality ratio $f$ [Eq. (9)] (open red circles) vs. time for eight of the countries most affected by the COVID-19 pandemic. Arrows indicate the dates when each country started social-distancing measures. For the USA, the date shown is the date which New York, an early epicenter of cases, enacted measures.

The examples in Fig. 2 also indicate large discrepancies between the fatality ratios of different countries. Significant discrepancies (e.g., between Italy or USA and South Korea) remained even after lockdown orders were issued (which otherwise had beneficial effects). Possibly, the observed differences are more likely related to criteria used in admitting patients into the hospital systems than to the actual fatality rate for the same kind of population. In other words, while South Korea was likely testing not only persons that showed symptoms but also persons the infected people interacted with (whether symptomatic or not), Italy, due to the magnitude and surprising evolution of their epidemic, had to resort to testing people with severe symptoms only. In light of this interpretation, and based on the data from South Korea, we notice in passing that the actual mortality rate for COVID-19 is 3% or less.

To elucidate the cause of the temporal changes within $f(t)$ [Eq. (9)] for each country and the differences seen from country to country, we plotted separately the death and recovery rates vs. time (see Fig. 3). The death rates were between 0.001 days$^{-1}$ and 0.005 days$^{-1}$ for the USA, China, and Germany, while for France the upper limit was over five times higher. China and South Korea managed to keep death rates low from the start of the outbreak, most likely due to their respectively meticulous responses, while the death rates increased initially in other nations. As with the fatality ratios (Fig.2), the death rates for each country begin to decline after implementation of social-distancing measures and a brief adjustment period, although for France and Brazil the delay was comparatively longer. The decline in death rates may be due to the response of the healthcare system which may have been initially overwhelmed, then finally adapted to the situation. The height of the peak, as well as the rate of decline to steady-state may serve as a measure of effectiveness of the various containment measures and the healthcare capacity of a nation. For large nations such as the USA, where the epidemic did not spread synchronously, the state governments did not enact social-distancing measures at the same time, and data collection protocols may have varied between states, the death rate experienced a slow growth and decline, yet it did not achieve a high peak. Overall, death rates as low as 0.0001 days$^{-1}$ to 0.001 days$^{-1}$ may be maintained as shown by South Korea and also seen in all the countries when the healthcare systems were no longer overwhelmed. This suggests that the COVID-19 pandemic may become less deadly if aggressive testing and contact tracing are enacted, coupled with better preparation of the hospital systems.

The recovery rates (also shown in Fig. 3) yield additional information. The variation in recovery rate ranges between nations may be related to the overall health of the populations, the quality of treatment of each respective healthcare system, as well as varying criteria as to when an individual's recovery is officially reported. For each nation (after the initial transient period), the recovery rates oscillate around a central value, which is seen to change over the course of the epidemic as government and healthcare systems adjust. For Italy, Germany, Iran, and China the value around which the recovery rate oscillates is seen to increase after the implementation of social-distancing protocols. For France, Brazil, and the USA, the immediate impact of social distancing on the recovery rate does not follow the trend of the other nations and may indicate that these nations either did not enact sufficient measures or that the populations and businesses did not sufficiently comply.

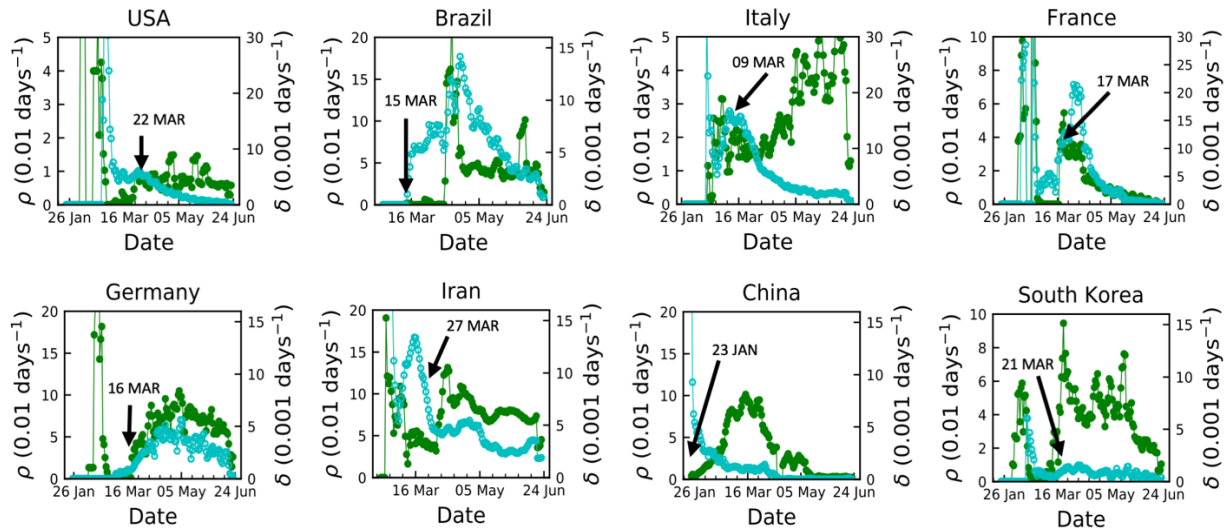

**FIG. 3.** Mortality (open blue circles) and recovery (closed green circles) rate constants vs. time, computed from Eqs. (4) and (5), from the start of testing for the same countries in Fig. 2.

It is also useful to monitor the average number of new infections caused by a person already infected at a given time. This is done using Eq. (11), and the results are summarized in Fig. 4. With the initial transient behavior excluded, each country shows an initial decline as testing capacity stabilizes. For the United States, a decline to a value of around 1.2 is seen followed by a disturbing steady increase toward a value of about 3.0. Germany, China, and South Korea achieved the fastest decline of the new infection ratios which finally hover for a significant period of time around 0.1 to 0.5. This method is sensitive to statistical noise, where short-term deviations from the typical behavior result in noise spikes in the curves, as seen around day 12 of April in France. Sub-unitary values of the new infection ratio indicate that more people are recovering/dying than are being infected and hence the epidemic is waning, and a prolonged period with a value of zero would indicate the presumptive end of the epidemic. A significant period of increase would indicate the re-emergence of infection hot spots, which may be the situation toward the end of June for China and South Korea, and thus that social-distancing measures were not having the desired effect and that these areas require further attention.

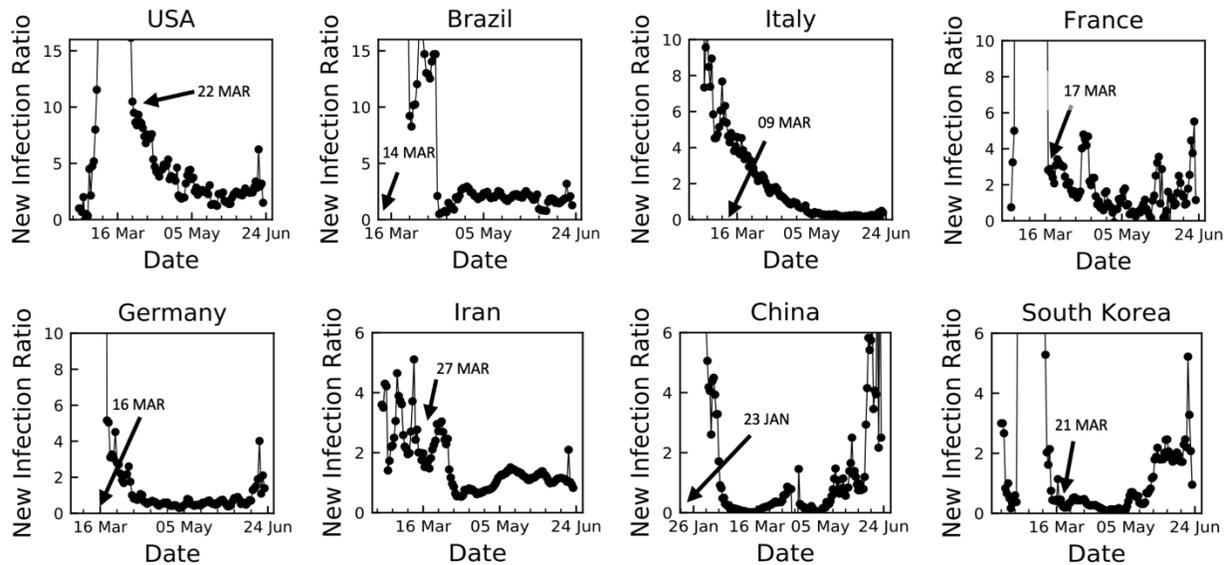

FIG. 4. New infection ratio [Eq. (11)] vs. time from the start of testing for each country shown in Figs. 2 and 3.